\documentclass{emulateapj}

\lefthead{Bekki \& Chiba}
\righthead{Dark impact}

\begin{document}

\title{Impact of dark matter subhalos on extended HI disks of galaxies: 
Possible formation of HI fine structures and stars}

\author{Kenji Bekki} 
\affil{
School of Physics, University of New South Wales, Sydney 2052, Australia}

\and

\author{Masashi Chiba}
\affil{
Astronomical Institute, Tohoku University, Sendai, 980-8578, Japan\\}

\begin{abstract}
Recent observations have discovered star formation activities  in 
the extreme outer regions of disk galaxies. 
However it remains unclear what physical mechanisms are responsible
for triggering star formation in such low-density gaseous environments
of galaxies.
In order to understand the origin of these outer star-forming
regions, we numerically investigate how the impact of 
dark matter subhalos orbiting a gas-rich disk galaxy embedded in a massive
dark matter halo influences the dynamical evolution of
outer HI gas disk of the galaxy. 
We find that if the masses of the subhalos ($M_{\rm sb}$) in a galaxy 
with an extended HI gas disk  are as large as 
$10^{-3} \times M_{\rm h}$, where $M_{\rm h}$ is the total mass of
the galaxy's dark halo,  
local fine structures 
can be formed in the extended HI disk.
We also find that the gas densities of some apparently filamentary structures
can exceed a 
threshold gas density for star formation and thus be likely 
to be  converted into
new stars in the outer part of the HI disk in some models 
with larger $M_{\rm sb}$. 
These results  thus imply
 that the impact of dark matter subhalos (``dark impact'')
can be important for better understanding the origin of recent star
formation discovered in the extreme outer regions of disk galaxies.
We also suggest that characteristic morphologies of local gaseous structures
formed by the dark impact can indirectly prove the existence of
dark matter subhalos in galaxies. We discuss the origin of 
giant HI holes observed in some gas-rich  
galaxies (e.g., NGC 6822) in the context of the dark impact.

\end{abstract}


\section{Introduction}

Star formation activities in the extreme outer regions
of gas-rich disk galaxies have recently come to be discussed
extensively not only  in the context of star formation laws in
low-density environments  of galaxies 
but also  in the context of formation and evolution of disk galaxies
(e.g., Ferguson et al. 1998a, b;
Leli\`evre \& Roy 2000; Cuillandre et al. 2001; 
Martin \& Kennicutt 2001; de Blok \& Walter 2003;
Gil de Paz et al. 2005; Thilker et al. 2005).
Recent observational studies of M83 by the {\it Galaxy Evolution Explore
(GALEX)} have discovered UV-bright stellar complexes associated
with filamentary HI structures in the extreme outer disk 
at $R \sim 4R_{\rm HII}$, where $R_{\rm HII}$ corresponds to
the radius where the majority of HII regions are detected
(Thilker et al. 2005).
Furthermore a number of small isolated HII regions have been
recently discovered at projected distance up to 
30 kpc from their nearest galaxy (e.g., NGC 1533)
in the {\it Survey for Ionization in Neutral Gas Galaxies (SINGG)}
(e.g., Meurer 2004; Ryan-Weber et al. 2003).

It has been discussed whether the observed properties of
these recent star formation activities  in the extreme outer HI regions
of disk galaxies can be understood in terms of
local gravitational instability within the HI disks
(e.g., Ferguson et al. 1998a;  Martin \& Kennicutt 2001).
Ferguson et al. (1998a) suggested that a simple picture
of the local gravitational instability can not explain
self-consistently the observed radial distributions
of HI gas and H$\alpha$ flux of star-forming regions.
Star formation (proven by H$\alpha$ regions)
in dwarf irregular galaxies 
(e.g., ESO 215-G?009)  
with extended HI disks
is observed to occur
in their very outer parts, where the gas densities
are well below a critical density of star formation 
(e.g., Warren et al. 2004). 
Thus it remains unclear what can control the star formation
activities in the extreme outer regions of disk galaxies.

Using numerical simulations of  galaxy formation  based on
the  cold dark matter (CDM) model,
Font et al. (2001)
demonstrated that dark matter subhalos predicted in
the CDM model
can play a minor role in the heating of the disk
owing to the very small number of subhalos approaching
to the solar radius of 8.5 kpc.
The sizes of HI disks of gas-rich galaxies are generally observed to be  
significantly larger than their optical disks 
with the sizes of $R_{\rm s}$ (Broeils \& van Woerden 1994)
and some  fraction
of low luminosity galaxies have HI gas envelopes extending out to 4--7
$R_{\rm s}$ (e.g., Hunter 1997).
No theoretical attempts have been made to investigate the dynamical
impact of dark matter subhalos (hereafter referred to as ``dark impact''
for convenience) on {\it the extended HI disks of galaxies},
though several numerical studies have already investigated
the influences of triaxial halos with figure rotation
and tidal galaxy interaction
on the evolution of the extended HI disks
(e.g., Theis 1999; Bekki \& Freeman 2002; Masset \&  Bureau 2003;
Bekki et al. 2005a, b).

The purpose of this Letter is to propose that the dynamical interaction
between dark matter subhalos and extended HI gas disks can
be important for better understanding the origin of recent
star formation observed in the extreme outer regions of
disk galaxies. By using hydrodynamical simulations of
the dark impact on galaxy-scale HI disks,
we show that local fine structures (e.g., filaments and holes)  can be
formed by the dark impact in the HI disks.
We discuss whether star formation can occur
in high-density regions of apparently  filamentary structures  formed by
the dark impact. 
We  suggest that characteristic fine structures formed by
the dark impact in a HI disk of a galaxy
{\it with apparently no interacting visible dwarfs close to the disk}
can indirectly prove the presence of
dark matter subhalos that frequently pass through the outer part
of the HI disk. 

\section{Model}

We investigate how the extended HI disk of a galaxy embedded
in a {\it fixed}  dark matter halo with the total mass of $M_{\rm h}$
is dynamically influenced by a {single subhalo} that is orbiting the galaxy
and represented by a {\it point mass.}
In order to elucidate the essence of the dynamical effects of the
dark impact more clearly, we adopt the above somewhat idealized 
model: We will describe the successive and cumulative
impact of {\it numerous} subhalos with a reasonable spatial distribution
within a {\it live} dark matter halo of a galaxy in our
forthcoming papers (Bekki \& Chiba 2005 in preparation).
We adopt the density distribution of the NFW
halo (Navarro, Frenk \& White 1996) suggested from CDM simulations
for the fixed dark matter halo:
 \begin{equation}
 {\rho}(r)=\frac{\rho_{s}}{(r/r_{\rm s})(1+r/r_{\rm s})^2},
 \end{equation}
where  $r$, $\rho_{s}$, and $r_{\rm s}$ are
the spherical radius,  the characteristic
 density of a dark halo,  and the scale
length of the halo, respectively. The $c$ parameter of the NFW halo
is set to be 10.0. The total baryonic mass ($M_{\rm b}$) 
of the galaxy is assumed
to be $0.1M_{\rm h}$.
Henceforth, all masses are measured in units of
$M_{\rm b}$ and  
distances in units of $r_{\rm s}$, unless otherwise specified.
Velocity and time are measured 
in units of $v$ = $ (GM_{\rm b}/r_{\rm s})^{1/2}$ and
$t_{\rm dyn}$ = $(r_{\rm s}^{3}/GM_{\rm b})^{1/2}$, respectively,
where $G$ is the gravitational constant and assumed to be 1.0
in the present study.
If we adopt $M_{\rm b}$ = 6.0 $\times$ $10^{10}$ $ \rm M_{\odot}$ and
$r_{\rm s}$ = 10.5\,kpc as fiducial values, then $v$ = 1.57 $\times$
$10^{2}$\,km\,s$^{-1}$  and  $t_{\rm dyn}$ = 6.55 $\times$ $10^{7}$ yr.

The gas disk 
with the total  mass of $M_{\rm g}$
is composed of $10^5$ SPH (Smoothed Particle Hydrodynamics)
particles and 
an isothermal equation of state with the sound speed of
$0.02 v$ is used for the gas.
The gas is assumed to have an uniform radial density distribution 
(rather than an exponential one) and distributed for  
$2r_{\rm s} \le R \le 4r_{\rm s}$,
because we intend to investigate the influence of the dark impact
on the dynamical evolution of the outer gas disk.
A reasonable dynamical model of the Galaxy embedded in 
the NFW halo has $r_{\rm s} \sim 0.6 R_{\rm s}$, where $ R_{\rm s}$
is the stellar disk size (Bekki et al 2005c).
Therefore the adopted gaseous distribution
corresponds to that
for $1.2R_{\rm s} \le R \le 2.4R_{\rm s}$
and thus can be regarded as reasonable for investigating
gas dynamics of outer HI disks of galaxies.
The gas disk is  assumed to be influenced  only by its host dark  halo  
(not by the inner stellar disk component),
because our interests are on the evolution of the very outer part
of the gas disk, 
where the gravitational field of its dark matter halo dominates.

The mass of the subhalo ($M_{\rm sb}$) is set to be a free
parameter that
can control the strength of the dark impact
and ranges  from $10^{-4} M_{\rm h}$ to  $10^{-2} M_{\rm h}$.
Although the initial position (${\bf X}_{\rm sb}$) and
velocity (${\bf V}_{\rm sb}$) are set to be free parameters,
we show the results of the models with
${\bf X}_{\rm sb}$ = ($x$,$y$,$z$) = ($3r_{\rm s}$, 0, $0.5r_{\rm s}$)
and ${\bf V}_{\rm sb}$ = ($v_{\rm x}$,$v_{\rm y}$,$v_{\rm z}$) 
=(0, 0, $0.5V_{\rm c}$), where $V_{\rm c}$ is the circular
velocity at ${\bf X}_{\rm sb}$ derived for the NFW halo. 
The results of other models with different ${\bf X}_{\rm sb}$
and ${\bf V}_{\rm sb}$ will be described in our forthcoming
papers (Bekki \& Chiba 2005). We mainly show the results
of  the ``fiducial model'' with 
$M_{\rm g}=0.01M_{\rm h}$ and $M_{\rm sb}=0.01M_{\rm h}$,
which shows more clearly the roles of the dark impact
in the dynamical evolution of HI gas disks in the present study.
All the calculations related to the above  hydrodynamical evolution
have been carried out on the GRAPE board (Sugimoto et al. 1990)
at the Astronomical Data Analysis Center (ADAC)
at the National Astronomical Observatory of Japan.
The gravitational softening parameter 
in the GRAPE5-SPH code (Bekki \& Chiba 2005)
is  fixed at 0.019 in our units
and the time integration of
the equation of motion is performed by using the predict-corrector
method with a multiple time step scheme.

\section{Result}

Figure 1 describes how gaseous fine structures are formed 
by the dark impact in the extended gas disk for the fiducial
model with $M_{\rm sb}/M_{\rm h}=0.01$.
As the dark matter subhalo passes through the thin gas disk,
it tidally disturbs the disk in a moderately strong manner.
Owing to the small ratio of $M_{\rm sb}/M_{\rm h}=0.01$,
the tidal field of the subhalo is not strong enough
to trigger the formation of global, non-axisymmetric  
structures (e.g., spiral arms and bars) and warps.
The dark impact  however can form local fine structures
that look like  ``filaments'' in the $x$-$y$ projection
and ``chimneys''  in the $x$-$z$ projection at $T=1.1$.
As the subhalo approaches the gas disk,
it gravitationally attracts gaseous particles along its path.
The vertical distribution of gaseous  particles close to the subhalo
passing through the gas disk
follows the wake induced by  the subhalo.
The particles 
consequently appear to  get levitated (or lifted up)
during and after the passage of the subhalo through the gas disk.
Accordingly chimney-like structures can be clearly seen if the HI disk
is viewed from the edge-on.
These fine structures can be clearly seen in other models
with $M_{\rm sb}/M_{\rm h} > 0.001$
and thus regarded as  characteristics of
local gaseous structures formed by the dark impact,
though the details of their morphologies are appreciably different
with one another.
 
Figure 2 shows the two-dimensional (2D) distribution of 
the projected gaseous densities (${\mu}_{\rm g}$) of the
simulated kpc-scale fine structures shown in Figure 1.
Because of the tidal compression of gas by the dark impact,
some parts of the structures show ${\mu}_{\rm g}$ higher
than the threshold gas density for star formation 
(${\mu}_{\rm thres} \sim 3 {\rm M}_{\odot}$ ${\rm pc}^{-2}$;
Hunter et al. 1998). ${\mu}_{\rm g}$ can be as high as
$\sim  14 {\rm M}_{\odot} $ ${\rm pc}^{-2}$ and
about 4 \% of the local regions (i.e., 105 among 2500 cells)
in Figure 2 have ${\mu}_{\rm g} > {\mu}_{\rm thres}$.
Although we do not model star formation in the present simulations,
these results imply that star formation in
the extreme outer parts of  gaseous disks of galaxies is highly likely
to be triggered by the dark impact. 

Figure 3 shows how the strength of the dark impact
controls the time evolution of ${\mu}_{\rm g}$ of outer 
gaseous disks of galaxies. 
The larger values of $M_{\rm sb}/M_{\rm h}$ mean that
the subhalos can give stronger dynamical impact  to the gas disks.
It is clear from Figure 3 that (1) the maximum density of 
local structures formed by the stronger dark impact
is higher
and (2) the number fraction of cells with 
${\mu}_{\rm g} > {\mu}_{\rm thres}$ is larger for the
stronger dark impact.
These results imply that star formation can occur
in the wider regions of extended HI disks of galaxies
when the HI disks are more strongly influenced by
the dark matter subhalos.

The long-term evolution of
local fine structures formed by the dark impact
should be investigated and described 
in our future works with models of star formation,
because star formation and its feedback 
(e.g., supernovae explosion) are highly likely  to  influence 
the evolution of the structures.
However it would be instructive to describe
some results possibly characteristic of the long-term evolution 
derived in the present models without star formation.
Figure 4 shows the 2D distribution of  ${\mu}_{\rm g}$
in the model with $M_{\rm sb}/M_{\rm h}=0.001$ at $T=6$.
As shown in this figure,
a kpc-scale hole surrounded by appreciably higher density regions
can be finally formed as a result of the dark impact.

This result implies that without energetic thermal and kinematic
feedback from massive OB stars and supernovae, giant HI 
holes can be formed in extended HI disks of galaxies by
the dark impact. Thus the dark impact provides a new
mechanism for the formation of giant HI holes observed
in gas-rich dwarf irregular galaxies (e.g., LMC).
The observed giant HI hole in NGC 6822  (e.g., de Blok \& Walater 2003)
might well be formed by dynamical impact of a very faint companion dwarf 
(observed as  ``NW cloud'')  on the HI disk: The physical mechanism
of the HI hole formation can be  essentially the same as the dark impact.
Thus the HI hole in the NGC 6822 system 
implies the viability of the dark impact scenario of  HI hole formation.

We confirm that the HI fine structures can be formed 
by the dark impact in models
with different $M_{\rm g}$ ($<0.05 M_{\rm h}$),
though the details of HI morphologies depend on $M_{\rm g}$ and
initial orbits of subhalos (i.e., ${\bf V}_{\rm sb}$).
The more remarkable chimney-like structures can be formed
in the models with higher inclination  of the orbits
of subhalos.  Apparently  filamentary structures with higher gas densities
can be formed by the dark impact in the model with 
a higher initial gas density of $M_{\rm g} = 0.02 M_{\rm h}$.

\section{Discussions and conclusions}

Although the present study has suggested 
that the dark impact can be responsible for
star formation in the outer HI gas disks of galaxies,
it remains unclear what roles the dark impact has
in the evolution of HI disks {\it within stellar 
disks of galaxies.}
Although the possibility of dark matter subhalos
approaching galactic stellar disks is low (Font et al. 2001),
we here suggest the following two possible roles of
the dark impact.
One is that the unique local structures of OB stars, young clusters,
and super-associations, such as the galactic belt and the Gould
belt in the Galaxy (e.g., Stothers \& Frogel 1974),
can result from the dark impact. The other is that
giant HI holes without bright optical stellar counterparts
(e.g., star clusters responsible for supernovae explosion 
that for giant HI holes)
observed in some low-luminosity galaxies 
can be due to the dark impact. 
These speculative suggestions need to be investigated in
a quantitatively way by our future high-resolution  simulations on
the dark impact on inner HI gas disks of galaxies.

The present study has shown that the previous passages of
dark matter subhalos through extended HI disks of galaxies
can be imprinted on fine structures (e.g., filaments and holes)
in the HI disks.
The present model also predicts that if a galaxy-scale halo
with an extended HI gas disk  has
$\sim 500$ subhalos,  $\sim 8$ HI fine structures can
be formed by dark impact for every 1 Myr. 
Recent HI observations have revealed that some fraction
of galaxies have very extended HI disks 
(e.g., Hunter 1997; Warren et al. 2004),
though the total number of galaxies whose  outer HI 
structures have been extensively investigated 
are  very small.
Accordingly we suggest that if galaxy halos are composed
of numerous subhalos, as the CDM model predicts,
future high-resolution HI observations  on
the fine structures of extended gas disks  for
a statistically significant number of galaxies
can {\it indirectly} probe the existence of the subhalos
and thereby provide some constraints on possible spatial distributions
and kinematics of the subhalos.
We also suggest that HI gas disks with apparently no interacting
(visible) dwarf galaxies  close to
the disks would be the best observational targets for proving
subhalos in galaxies.

\acknowledgments
We are  grateful to the  referee  Fabio Governato for valuable comments,
which contribute to improve the present paper.
K.B. acknowledges the financial support of the Australian Research 
Council throughout the course of this work.
The numerical simulations reported here were carried out on GRAPE
systems kindly made available by the Astronomical Data Analysis
Center (ADAC) at National Astronomical Observatory of Japan (NAOJ).


\clearpage

\begin{figure}
\epsscale{.50}
\plotone{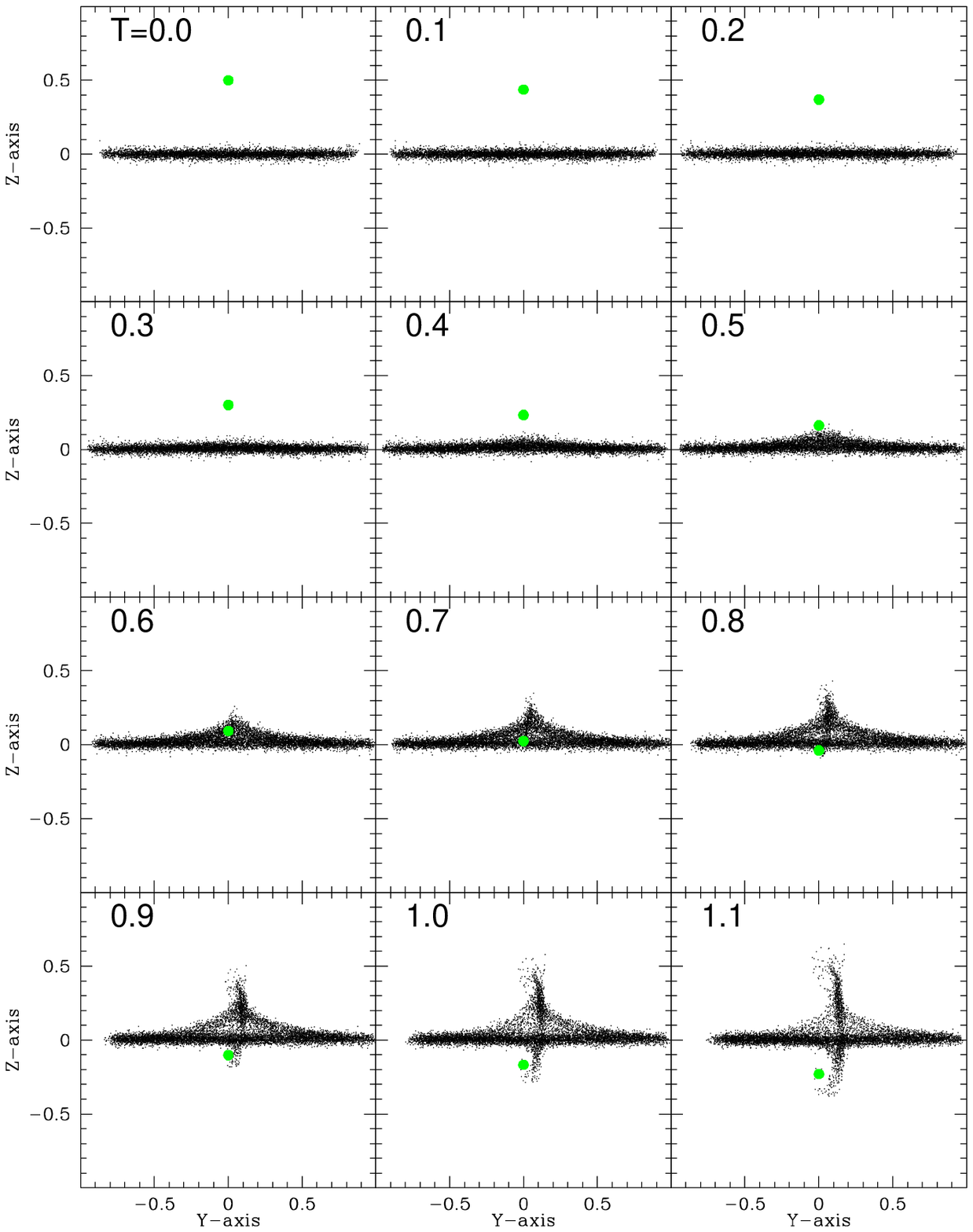}
\caption{
Gaseous distributions projected onto the $x$-$z$ plane
for the fiducial
model with $M_{\rm sb}/M_{\rm h}=0.01$. 
The coordinate of the center of each frame
is ($x$,$y$,$z$)=($x_{\rm sb}$, $y_{\rm sb}$, 0),
where $x_{\rm sb}$ and $y_{\rm sb}$ are the coordinate of
the subhalo at each time step, so that
the time evolution of the gas distribution influenced by the dark impact
can be more clearly seen.
The time $T$ 
(in the simulation units) shown
in the upper left corner of each panel represents time
that has elapsed since the simulation starts.
The position of the dark matter subhalo  is shown 
by a green big dot. For clarity, only gas particles within
$r_{\rm s}$ (=1) of the subhalo are shown. 
\label{fig-1}}
\end{figure}

\clearpage

\begin{figure}
\epsscale{.80}
\plotone{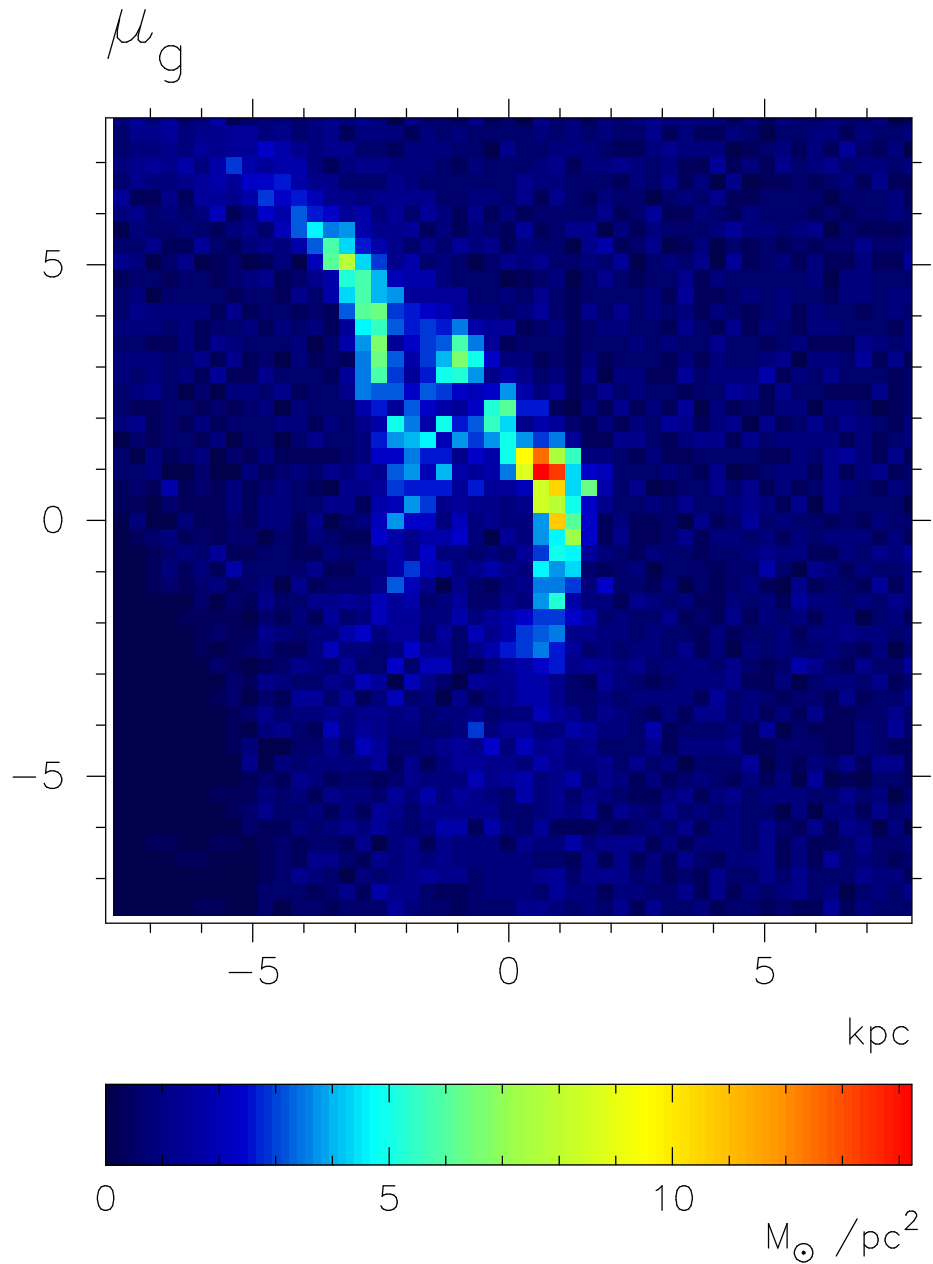}
\caption{
The two-dimensional (2D) gas  density distributions (${\mu}_{\rm g}$
in units of ${\rm M}_{\odot}$ pc$^{-2}$) projected onto
the $x$-$y$ plane for the simulated 
fine structures formed by the dark impact in the fiducial model
at $T=2$.
For clarity, the high-density gaseous regions are set to coincide
with the center of the flame.
The 2D map consists of 2500 cells with the cell size of 0.315 kpc
and ${\mu}_{\rm g}$ is estimated for each cell.
Note that some parts of the apparently filamentary structures exceed
the threshold gas density 
for star formation
(${\mu}_{\rm thres} \sim 3 {\rm M}_{\odot}$ pc$^{-2}$; Hunter et al. 1998).
\label{fig-2}}
\end{figure}

\clearpage

\begin{figure}
\epsscale{1.0}
\plotone{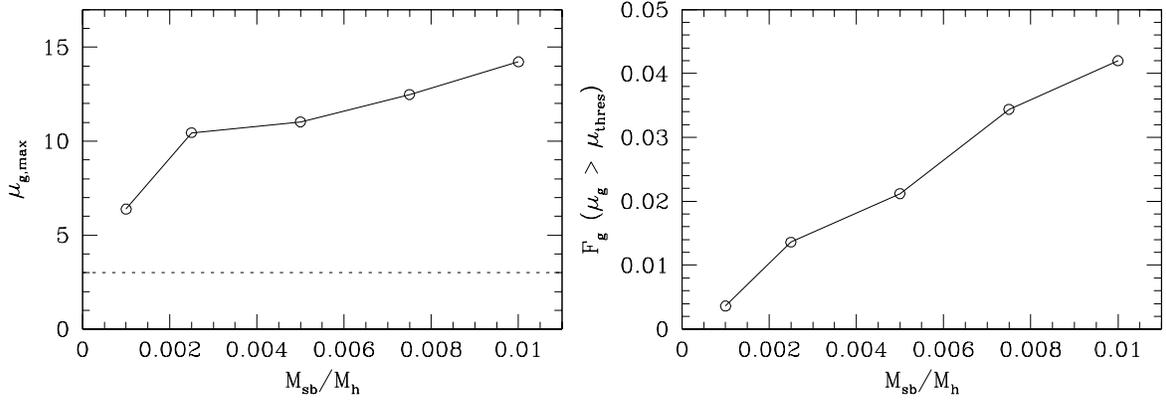}
\caption{
The dependences of the maximum projected gas density (${\mu}_{\rm g,max}$)
of the simulated fine structures (left) and the number fraction
of cells ($F_{\rm g}$) with ${\mu}_{\rm g} > {\mu}_{\rm thres}$ (right)
on $M_{\rm sb}/M_{\rm h}$.
${\mu}_{\rm thres}$ ($\sim 3 {\rm M}_{\odot}$ pc$^{-2}$) shown
by a dotted line in the left panel
is the observed threshold gas density for star formation. 
Note that the stronger the dark impact is (i.e., the larger
$M_{\rm sb}/M_{\rm h}$ is), the larger ${\mu}_{\rm g,max}$
and  $F_{\rm g}$ are. 
\label{fig-3}}
\end{figure}

\clearpage

\begin{figure}
\epsscale{.80}
\plotone{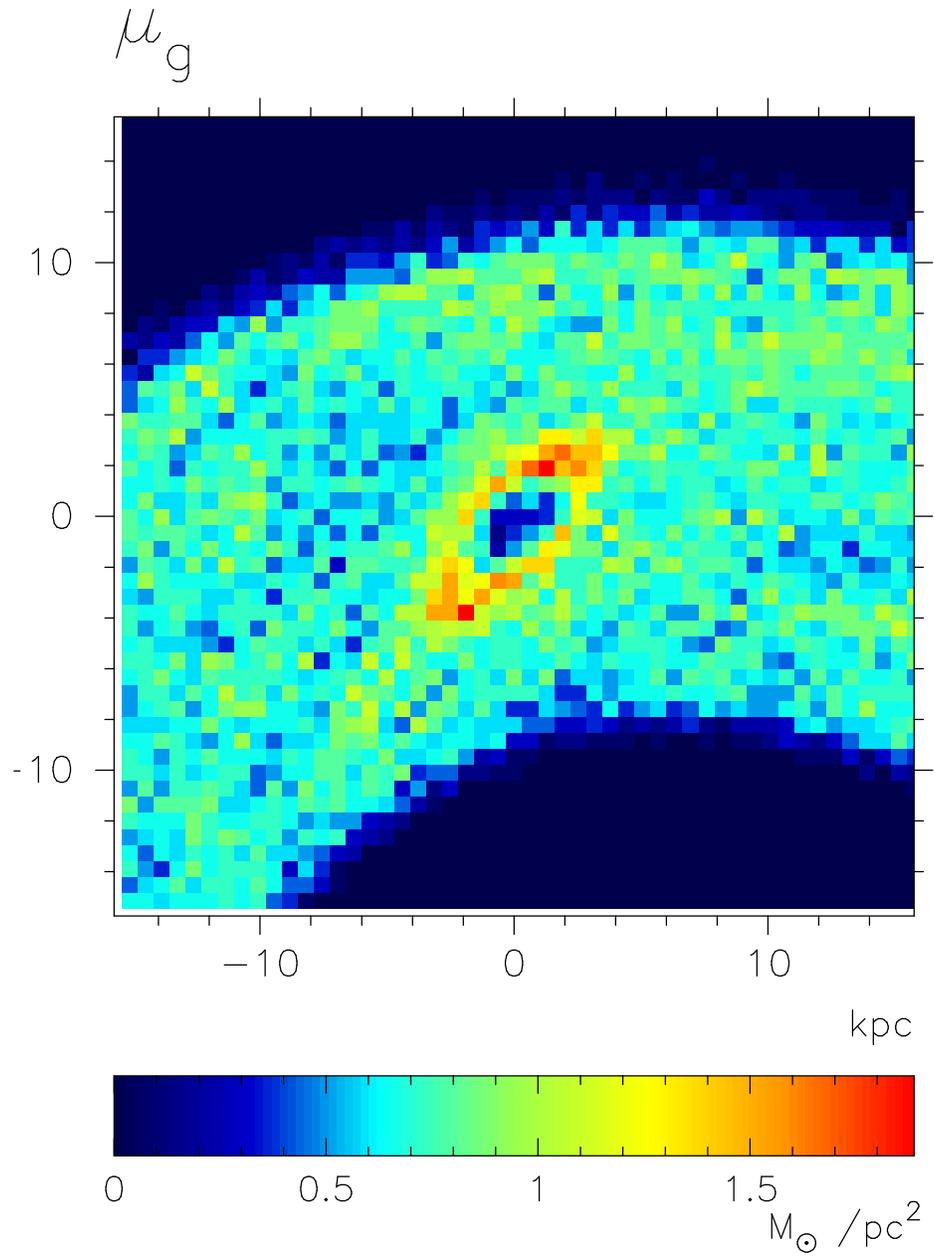}
\caption{
The same as Figure 2 but for the model with $M_{\rm sb}/M_{\rm h}=0.001$
and $T=6$. Note than a hole-like structure can be seen in the center
of this frame.
\label{fig-4}}
\end{figure}

\end{document}